\documentclass[prl,floatfix,twocolumn,showpacs]{revtex4}

\usepackage{graphicx}%
\usepackage{dcolumn}
\usepackage{amsmath}
\begin{document}

\preprint{cond-mat/none so far}

\title[Short Title]{Emergence of self-sustained patterns in small-world 
excitable media}
\author{Sitabhra Sinha$^1$}
\author{Jari Saram\"{a}ki$^2$}
\author{Kimmo Kaski$^2$}
\affiliation{%
$^1$The Institute of Mathematical Sciences, C.I.T. Campus, Taramani, Chennai - 600 113 India\\
$^2$Laboratory of Computational Engineering, Helsinki University of Technology, P.O. Box 9203, FIN-02015 HUT, Finland
}%

\date{\today}
\begin{abstract}
%
%

Motivated by recent observations that long-range connections (LRC) play a role 
in various 
brain phenomena,
we have 
observed two distinct dynamical transitions in the activity
of excitable media 
where waves propagate both
between neighboring regions and through LRC. 
When LRC density $p$ is low, single or multiple spiral waves 
are seen to emerge and cover the entire system. This state is self-sustaining and robust 
against perturbations. At $p=p_c^l$ the spirals are suppressed and there 
is a transition 
to a spatially homogeneous, temporally periodic state.
Finally, above $p=p_c^u$, activity ceases after a brief transient.
%
\end{abstract}
\maketitle

Pattern formation in excitable media has recently been a very exciting area of
research, not least because it is observed in a wide
variety of natural systems, ranging from spiral waves in mammalian brain 
\cite{Hua04} to chemical systems such as the Belousov-Zhabotinsky reaction
\cite{Zai70}. The functional role of such patterns in biological systems 
makes it imperative to understand better the conditions in which 
they can spontaneously emerge. For example, these patterns have been implicated 
in the genesis of life-threatening arrhythmias in the heart \cite{Gra98}, while 
in the brain they have been thought to provide a spatial framework for 
cortical oscillations \cite{Hua04}. Until now, work has mostly focussed on 
generating patterns in excitable media by using stochastic stimulation 
\cite{Gar99}. In this case, however, the activity is not really self-sustained 
as the noise is akin to external intervention necessary for the initiation 
and persistence of spiral waves \cite{Jun95,Hou02}. If such patterns are to be 
seen as spontaneously emerging from arbitrary initial conditions, then the 
pattern formation should be an outcome of the internal structure of the system
alone. Furthermore, small variations in this 
structure may result
in transitions between different dynamical regimes characterized by
distinct spatiotemporal patterns.

Here we have considered excitable systems which have a regular topology with 
cells communicating only with nearest neighbors, but where there are a few 
random long-range connections, linking spatially distant cells. Such ``small-world'' 
topologies have been observed in a large number of real world systems \cite{Wat98}, 
and have also been associated with self-sustained activity in a chain of model neurons 
\cite{Rox04} as well as periodic epidemic patterns in disease spreading \cite{Kup01}. 
However,
while it has been shown that such long-range connections do play a role in 
maintaining 
spiral waves that already exist \cite{He02}, so far there have been very few
attempts at showing the emergence of, and transitions between, different types 
of self-sustained patterns as a result of such topology.
The observation of \emph{spontaneous} pattern formation in
natural systems where sparse long-range connections coexist with
fairly regular underlying connection topology, 
like in the brain \cite{Har98}, 
points towards intriguing possibilities for the functional role of these 
connections. In this Letter we have investigated a generic model of excitable
media with increasing density of random long-range connections, and
we report the existence of two qualitatively different regimes of 
self-sustained pattern formation. The correspondence of the observed patterns
with those observed in nature, e.g., epileptic bursts and seizures
\cite{Net04}, as well as their dependence on the topological structure of 
connections, makes these results highly relevant in our view. 

The model we consider here consists of a two-dimensional array of $N \times N$
excitable cells coupled diffusively,
\begin{displaymath}
x^{i,j}_{t+1} = (1 - D) f ( x^{i,j}_t, y^{i,j}_t) \\ + \frac{D}{4} \sum_{q=\pm 1} 
f (x^{i+q,j+q}_t, y^{i+q,j+q}_t ).
\end{displaymath}
Here $D$ is the diffusion coefficient and the dynamics of individual cells are 
described by a pair of variables ${ x_t, y_t }$, 
evolving according to a discrete-time 
model of generic excitable media \cite{Chi95}:
\begin{eqnarray}
x_{t+1} & = & f (x_t, y_t ) =  x_t^2 e^{(y_t - x_t)} + k, \nonumber \\
y_{t+1} & = & g (x_t, y_t ) =  a y_t - b x_t + c, \nonumber 
\end{eqnarray}
where we have fixed the parameters as
$a =0.89, b = 0.6, c = 0.28$, and $k = 0.02$. When this system is 
excited with a suprathreshold stimulation, the fast variable $x$
shows an abrupt increase. This triggers changes in the
slow variable $y$, such that the state of the cell
is gradually brought down to that of the resting state. 
Once
excited, the cell remains impervious to stimulation up to a
refractory period, the duration of which is governed by the parameter 
$a$. Neighboring cells communicate excitation to each other with 
a strength proportional to 
$D$, chosen $D=0.2$ 
for most of our simulations. 
In addition
to the diffusive coupling, we introduce long-range connections such that 
each cell receives a connection from a randomly chosen
cell with probability $p$. The strength of this connection is chosen to 
be the same as that of the nearest neighbors, i.e., $D/4$. These random
long-range links can be either quenched (i.e., chosen initially and
kept fixed for the duration of the simulation), or annealed
(i.e., randomly created at each time step).
While the results reported below are for annealed random links, we could observe no
qualitative difference between these two cases. Note that we have used both
periodic and absorbing boundary conditions for the system, and observed no
significant difference in the results. All the following results were obtained
for absorbing boundary conditions.

In the simulations, the observed patterns formed in the system for varying 
values of 
the shortcut density $p$ can, generally speaking, be divided into three 
categories. First,
below the lower critical probability, i.e., $0<p<p_c^l$,  the state of the system after 
an initial transient period is characterized by self-sustaining single or multiple 
spiral waves covering the entire system. Second, at $p=p_c^l$, the spiral wave mode 
is suppressed and the system undergoes a transition to a regime in which a large 
fraction of the system gets simultaneously active, and then refractory, in a periodic 
manner. Third, when the value of $p$ is increased above a system-size-dependent upper 
critical probability $p_c^u$, the self-sustained activity ceases and the system falls 
into the absorbing state where $x^{i,j}=0, \forall i,j$.

\begin{figure}
\includegraphics[width=0.75\linewidth]{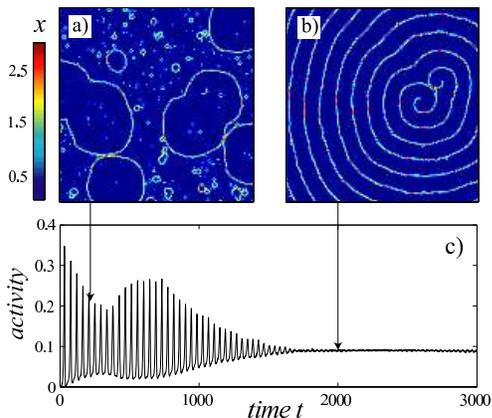}
\caption{Emergence of spiral waves, recorded in a simulation of a 
$N^2 =128\times128$ system, with shortcut probability $p=0.25$. 
After initialization, the dynamics of the system can be characterized
by circular waves. At around $t \sim 1,500$ time steps, a spiral wave is 
spontaneously created and 
subsequently takes over the dynamics. Panels (a) and (b) show the
state of the system at different times in terms of the fast variable $x$.
Colors indicate excitation level of the cells. (c) Time series
of the average activity, i.e. the fraction of cells with $x>0.9$. 
}\label{fig:circular_to_spiral}
\end{figure}

\begin{figure}[b]
\includegraphics*[width=0.95\linewidth]{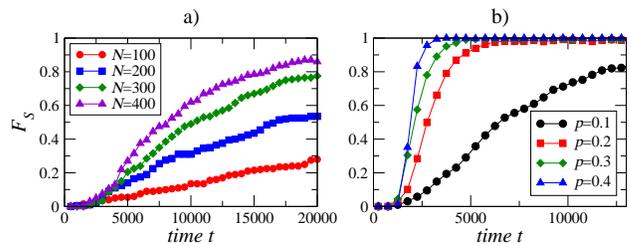}
\caption{Fraction $F_S$ of systems with spiral waves in 400 runs as function of 
time $t$: (a) fixed shortcut probability $p=0.05$ with varying system size, 
(b) fixed system size $N^2 = 300\times 300$ with varying shortcut probability $p$.}
\label{fig:timeseries}
\end{figure}

In the beginning of each simulation run the system was initialized such that $x=1$ for a 
small number of cells, and $x=0$ for the rest. 
For small shortcut probabilities $p \ll p_c^l$, upon starting the simulation,
multiple coexisting circular excitation waves were seen to emerge 
(see Fig.~\ref{fig:circular_to_spiral}a), to be later taken over by spiral
waves (Fig.~\ref{fig:circular_to_spiral}b). The activity time series in 
Fig.~\ref{fig:circular_to_spiral}c, displaying the fraction of cells where 
$x>0.9$, shows a high frequency periodicity corresponding to the refractory period; 
then, the periodicity disappears as the system settles into the spiral wave mode. 
For the time series preceding the onset of the spiral wave, we typically
observe slow modulations of the envelope of the periodic oscillations, which
arise from the interaction between the waves as well as from the shortcut-induced 
excitations. Power spectral densities of such series were observed to show
a power-law like decay, indicating the presence of $1/f$-noise. 

\begin{figure}[t]
\includegraphics[width=0.9\linewidth]{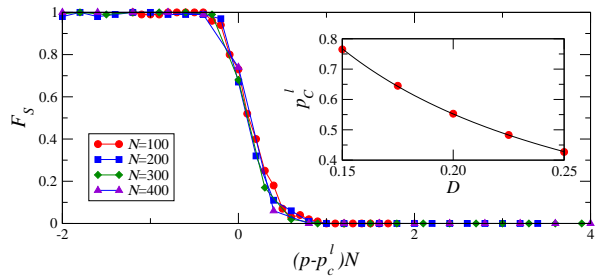}
\caption{ 
The fraction of spiral configurations $F_S$ as a function 
of $\left(p-p_c^l\right)$ normalized by 
system size $N$, for shortcut probabilities $p$ around $p_c^l\approx0.553$.
$F_S$ is calculated at $t=20,000$ time steps, averaged over 400 runs.
Inset: Dependence of the critical value $p_c^l$ on diffusion constant $D$.
The circles are simulation results for $N = 200$ while the curve shows fitting 
with $p_c^l \sim D^{-1.14}$.}\label{fig:pc}
\end{figure}

The spiral waves were observed to be primarily created by a shortcut-induced 
excitation occurring in the refractory ``shadow'' of a circular wave front, 
sparking a semi-circular wave whose transmission is partially hindered by the shadow. 
We verified that indeed spiral waves could be triggered in this fashion by 
using externally applied signal to stimulate an appropriate 
point in this region and then observing the resultant patterns.
Once a spiral wave is created, it will eventually take over the dynamics of the system,
as the successive excitation wavefronts occur with the highest frequency
compared to all other excitations which will then be swept away.
Evidently, the probability of a spiral wave creation per unit time 
increases with the system size $N^2$ and the shortcut probability $p$. 
Thus for long times and large system sizes, excitation via 
random shortcuts will eventually always result in the formation of 
spiral waves. This is corroborated by Fig.~\ref{fig:timeseries}, where the 
fraction of spiral wave configurations in 400 simulation runs is shown as 
function of time $t$, for varying system sizes at $p=0.05$ (panel {\em a}) and 
for fixed system size but increasing values of $p$ (panel {\em b}). 
The spatial structure of the spiral waves, i.e. an excitation front followed 
by a refractory shadow, makes them very robust against perturbations. 
For example, if the state of the system is frozen, and the state of large areas 
(say, upto a quarter of the total area) or every second cell are set to $x=0$, 
the spiral wave mode is quickly recovered once the simulation is restarted. 

\begin{figure}[t]
\includegraphics[width=0.98\linewidth]{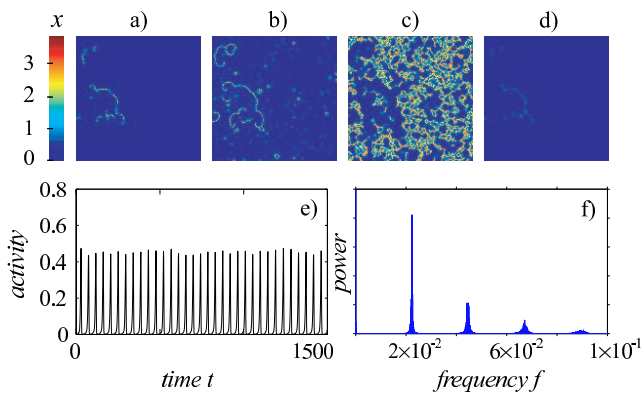}
\caption{The periodic regime, recorded in a simulation of a $N^2 =128
\times 128$ system, with shortcut probability $p=0.6$. a)-d): Snapshots
of the state of the system taken at intervals of $\Delta t=10$ steps. e): 
Time series of the activity (see Fig.~1) and (f): the corresponding power
spectrum.
The main peak is at $f_0 \approx 0.022$, corresponding to a wavelength of 
$\approx 43$ time steps; other peaks are harmonics at integer 
multiples of $f_0$.
}\label{fig:upc_snap}
\end{figure}

However, at high enough values of $p$, 
the shortcut-induced excitations become too numerous for sustaining the spiral 
wave dynamics. As almost every point is liable to be excited with a frequency
proportional to its refractory period, the spirals become unstable and 
we see a transition to a new regime at $p=p_c^l \approx 0.553$
(Fig.~\ref{fig:pc}). 
This value of $p_c$ was found to be independent of the system size $N^2$. 
Here, the spatial pattern becomes more homogeneous, as a large fraction of 
the system becomes simultaneously active and subsequently decays to a refractory 
state. However, some cells not participating in this wave of excitation carry 
on the 
activity to the next cycle, where it
again spreads through 
almost the entire system. This results in a remarkably periodic behavior of 
the system in time, with a large fraction of cells being recurrently active 
with a period close to the refractory period of the cells 
(see Fig.~\ref{fig:upc_snap} (e)-(f)). Often, small spiral-like waves were also 
observed (Fig.~\ref{fig:upc_snap} (a)-(d)), but these were short-lived and 
spatial correlations could not be maintained.

\begin{figure}[t]
\includegraphics[width=0.9\linewidth]{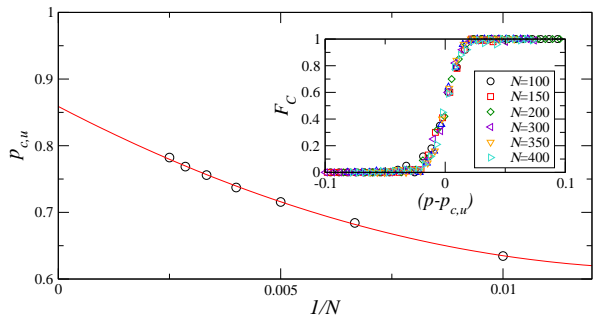}
\caption{Inset: Fraction of configurations $F_C$ where activity has ceased  at 
$t=20,000$ time steps in 100 simulation runs as function of $\left(p-p_c^u\right)$.
Main: The upper critical shortcut probability $p_c^u$ as function of inverse
system dimension $1/N$ ($\circ$). The solid line displays a fitted
quadratic function, where $p_c^u\left(N \rightarrow \infty \right)\approx 0.86$.
}\label{fig:upper_pc}
\end{figure}

Finally, when $p$ is increased still further, the very large number of shortcut 
connections guarantees almost simultaneous spread of excitation to nearly all 
cells. As a result, the dynamics of the system tends to ``burn out'' such 
that after a transient, almost all cells become refractory and not 
enough susceptible cells are left to sustain the excitation. We observed that 
once a system-size dependent value $p=p_c^u\left(N\right)$ is approached, 
the probability of reaching the absorbing state $\left(x^{i,j}=0, \forall i,j\right)$ 
grows rapidly (see Fig.~\ref{fig:upper_pc}) such that at high enough values of $p$, 
the system is never seen to sustain its activity. The limiting behavior 
for $N\rightarrow \infty$ was investigated by plotting $p_c^u$ against $1/N$ and 
extrapolating its value for $1/N\rightarrow 0$. Fitting a quadratic function 
yielded $p_c^u\rightarrow 0.86$, hence it appears that the regime where activity 
is sustained has for all system sizes an upper limit for the shortcut density $p$. 
Note that this can be viewed as an approximation only, because there is no 
\emph{a priori} reason to assume any particular form for the dependence
of $p_c^u$ on $N$. 

We have also carried out simulations with disorder in the parameters
describing the properties of individual cells. For instance, we have 
made the parameter $a$ that controls the 
refractory period a quenched random variable ranging
over a small interval. 
We find that there 
is a tendency for greater fragmentation of waves with disorder, and corresponding 
increase in the number of coexisting spiral waves, but otherwise no remarkable 
changes. The robustness of our results in the presence of disorder in 
the individual cellular properties underlines the relevance of this study to 
real-world systems, where 
cells are unlikely to have uniform 
properties. In addition, we have looked at the effect of the diffusion 
constant $D$.
For higher $D$ the wave propagates much faster, so that, in a fixed amount of time, 
the system will initiate excitation through many more long-range connections
than the system with lower $D$. 
Therefore, the higher $D$ system will be 
equivalent to a lower $D$ system with a larger number of long-range 
connections, 
i.e., higher $p$ (Fig.~\ref{fig:pc}, inset).

The above results carry potential relevance to all natural systems that are
excitable and have sparse, long-range connections. Experiments carried out
in the excitable BZ reaction system with nonlocal coupling \cite{Tinsley05} 
have exhibited some of the features described above. For example, the
coexistence of transient activity and sustained synchronized oscillations 
that we observe close to $p_c^u$ has been reported in the above system,
and can now be understood in terms of the model introduced here.
Recently, there has also been a lot of research activity on the role of
non-trivial network topology on brain function (e.g., Ref. \cite{Egu05}).
In particular, studies show a connection between the existence 
of small-world topology and epilepsy \cite{Net04,Per05}. 
This is one of the areas where our results can have a possible explanatory role.
In the brain, glial cells form a 
matrix of regular topology with cells communicating between their nearest
neighbors through calcium waves. Neurons are embedded on this regular
structure and are capable of creating long-range links between spatially
distant regions. It is now known that neurons and glial cells can communicate
with each other through calcium waves \cite{Cha94}. Therefore, the aggregate
system of neurons and glial cells can be seen as a small-world network
of excitable cells. Here the neurons, that are outnumbered by glial cells
approximately by one order of magnitude, 
form the sparse, long-range connections.
Recently, the role of neural-glial communication in epilepsy has been investigated
(e.g., Ref. \cite{Nad03}), but the observation of spiral
intercellular calcium waves in the hippocampus \cite{Har98}, and the 
similarity of the patterns seen in our model with the 
observed features of epileptic seizures and bursts, makes it especially 
appealing to postulate the role of small-world topology in the generation of 
epilepsy. Experimental verification of this suggested scenario can be
performed through calcium imaging in glial-neuronal co-culture systems. 

The relationship between the fraction of long-range connections and 
normal brain function is indicated by experimental evidence not
just for epilepsy but for various other mental disorders as well.
A recent study has found that the brains of patients with clinically 
diagnosed schizophrenia, depression or bipolar disease have
lower glia to neuron ratio compared to normal subjects \cite{Brauch06}.
Therefore, understanding the dynamical ramifications of increasing
shortcuts in an excitable media can motivate experiments that have 
the potential significance of
aiding clinical breakthroughs in treating a whole class of mental
disorders. This is also connected with the question of 
the evolutionary significance for increasing glia to neuron ratio 
with brain size \cite{Reichenbach89}. It is known that excitable media of 
larger dimensions are more likely to exhibit spiral waves \cite{Sinha01}. 
Therefore, 
decreasing the fraction of neurons (and therefore, long-range shortcuts) could 
be Nature's way of ensuring dynamical stability for neural activity.

To conclude, we have investigated spontaneous pattern formation in excitable
media with small-world connections. 
Our results show the creation of non-trivial
spatio-temporal patterns similar to those seen in many real-life systems.
These patterns
are formed through dynamics that is driven by the system's own internal architecture.
Most important of all, the system exhibits a non-trivial transition 
point at which the pattern goes from the spatial domain, with multiple
coexisting spiral waves where the global activity level remains more or
less uniform, to the temporal domain, where the global activity level shows
large oscillations as a large fraction of cells becomes simultaneously
active and then refractory, with a strict periodicity. The connection to 
biological phenomena, most importantly, to calcium waves in the brain, 
and the possibility of the functional role of small-world connections in 
epileptic seizures and bursts, is expected to make our study of special 
relevance.

\end{document}